\documentclass[conference]{IEEEtran}
\IEEEoverridecommandlockouts

\usepackage{fancyvrb}
\usepackage{xurl}
\usepackage{booktabs}
\usepackage{url}
\usepackage{amsmath,amsfonts}
\usepackage{algorithmic}
\usepackage{graphicx}
\usepackage{textcomp}
\usepackage{xspace}
\usepackage{subcaption}
\usepackage{tabularx}
\usepackage{amsmath}
\usepackage{balance}
\usepackage{framed}
\usepackage{chngpage}
\usepackage[many]{tcolorbox}
\usepackage{siunitx}
\usepackage[utf8]{inputenc}
\usepackage{fancybox}
\usepackage{booktabs}
\usepackage{multirow}
\usepackage{hyperref}
\usepackage[table,xcdraw]{xcolor}

\usepackage{graphicx}
\usepackage{subcaption}

\usepackage[htt]{hyphenat}
\usepackage{tikz}

\newcommand{\motivation}{\noindent\textbf{Motivation. }}
\newcommand{\approach}{\noindent\textbf{Approach. }}
\newcommand{\results}{\noindent\textbf{Results. }}

\newcommand{\rqa}{$RQ_1$}
\newcommand{\rqb}{$RQ_2$}
\newcommand{\rqc}{$RQ_3$}
\newcommand{\rqd}{$RQ_4$}
\newcommand{\rqaa}{Do VRT-related Reviews Get Merged More and Faster?}
\newcommand{\rqbb}{Are VRT-related Reviews Discussed More?}
\newcommand{\rqcc}{Do VRT-related Reviews Require More Complex Fixes?}
\newcommand{\rqdd}{What Problems Do Developers Discuss with VRT-Tests?}

\newcommand{\rqA}{\rqa: \rqaa}
\newcommand{\rqB}{\rqb: \rqbb}
\newcommand{\rqC}{\rqc: \rqcc}
\newcommand{\rqD}{\rqd: \rqdd}

\newcommand{\prs}{307\xspace} 

\newcommand{\exludepr}{7\xspace}
\newcommand{\prswithInopen}{314\xspace}
\newcommand{\projectsInclosed}{103\xspace}

\newcommand{\commentsInclosed}{373\xspace}
\newcommand{\visualPrs}{299\xspace}
\newcommand{\visualMergedPrs}{259\xspace}

\newcommand{\validVRTprs}{282\xspace}

\newcommand{\validVRTcomments}{344} 
\newcommand{\validVRTclassificationComment}{189}

\newcommand{\categoryAppearance}{Appearance\xspace}
\newcommand{\categoryLayout}{Layout\xspace}
\newcommand{\categoryColor}{Color\xspace}
\newcommand{\categoryText}{Text\xspace}
\newcommand{\categoryValue}{State\xspace}
\newcommand{\categoryPicture}{Image\xspace}
\newcommand{\categoryTest}{Test\xspace}

\newcommand{\sizeOfObjects}{Box Size Changed\xspace}
\newcommand{\newContents}{New Contents\xspace}
\newcommand{\disappearedContents}{Contents Disappeared\xspace}
\newcommand{\shapeOfObjects}{Box Shape Changed\xspace}
\newcommand{\disappearedObjects}{Box Disappeared\xspace}
\newcommand{\disappearedLines}{Border Disappeared\xspace}
\newcommand{\thicknessOfLine}{Wrong Border Thickness\xspace}
\newcommand{\newButton}{New Buttons\xspace}
\newcommand{\aNewOutlineAppear}{New Border Appeared\xspace}

\newcommand{\layoutShift}{Layout Shifted\xspace} 
\newcommand{\specificItems}{Specific Items are Mis-aligned\xspace}
\newcommand{\layoutChange}{Layout Changed\xspace}
\newcommand{\sizeOfPageHeader}{Size of Page/Header Changed\xspace}

\newcommand{\colorOfObjects}{Box Color Changed\xspace}
\newcommand{\colorOfTexts}{Text Color Changed\xspace}
\newcommand{\colorOfLines}{Border Color Changed\xspace}

\newcommand{\textModification}{Contents of Text Changed\xspace}
\newcommand{\someTextsDisappear}{Text Disappeared\xspace}
\newcommand{\sizeOfTexts}{Text Size Changed\xspace}
\newcommand{\someTextsAppear}{New Text\xspace}

\newcommand{\undefinedValue}{Undefined/Wrong State is Shown\xspace}
\newcommand{\defaultValue}{Default State is Changed\xspace}

\newcommand{\NoVisual}{No or Few Visual Differences Detected\xspace}
\newcommand{\fragile}{Fragile Snapshot\xspace}
\newcommand{\focusNotWork}{Focus Does Not Work\xspace}

\newcommand{\flackyTest}{Flaky Test\xspace}
\newcommand{\noChange}{No Changes\xspace}

\newcommand{\pictureAdjustment}{Image Adjusted\xspace}
\newcommand{\pictureReplacement}{Image Replaced\xspace}

\newtcolorbox{myRQbox}{
  colback=black!5!white, 
  colframe=black!75!white, 
  coltext=black,         
  fonttitle=\bfseries,
  boxrule=1pt,         
  arc=0mm,               
  boxsep=2pt,            
  left=2pt,              
  right=2pt,             
  top=3pt,               
  bottom=3pt,            
  breakable,             
}

\newcommand{\acceptRateVrt}{91.9}

\newcommand{\mergedVrtPR}{282}

\newcommand{\timeMergeMedianVrt}{4.5} 

\newcommand{\acceptRateVisual}{86.6}
\newcommand{\mergeVisualPR}{259}

\newcommand{\timeMergeMedianVisual}{1.2}

\newcommand{\commentMedianVrt}{10}
\newcommand{\changeFileVrt}{7}
\newcommand{\commitsVrt}{7}
\newcommand{\addedLineVrt}{129.5}
\newcommand{\deletedLineVrt}{40.5}

\newcommand{\commentMedianVisual}{1}
\newcommand{\changeFileVisual}{4}
\newcommand{\commitsVisual}{3}
\newcommand{\addedLineVisual}{52}
\newcommand{\deletedLineVisual}{9}

\pgfmathsetmacro{\calcAcceptDiff}{(\acceptRateVrt)/\acceptRateVisual}
\pgfmathsetmacro{\calcAcceptPer}{(\acceptRateVrt-\acceptRateVisual)/\acceptRateVrt*100}
\pgfmathsetmacro{\calcTimeDiff}{(\timeMergeMedianVrt)/\timeMergeMedianVisual}
\pgfmathsetmacro{\calcCommentDiff}{(\commentMedianVrt)/\commentMedianVisual}

\newcommand{\commentEffect}{-0.809}
\newcommand{\commitCiff}{-0.348\xspace}
\newcommand{\deleteCiff}{-0.302\xspace}
\newcommand{\addedCiff}{-0.266\xspace}
\newcommand{\changeFileCiff}{-0.285\xspace}

\newcommand{\invalidcomment}{155\xspace}

\newcommand{\duplicate}{58\xspace}
\newcommand{\inaccessible}{17\xspace}
\newcommand{\chromaticerror}{10\xspace}
\newcommand{\lackofcontext}{9\xspace}
\newcommand{\newstory}{61\xspace}

\newcommand{\categoryAppearanceNum}{52}
\newcommand{\sizeOfObjectsNum}{18\xspace}
\newcommand{\newContentsNum}{10\xspace}
\newcommand{\disappearedContentsNum}{6\xspace}
\newcommand{\shapeOfObjectsNum}{6\xspace}
\newcommand{\disappearedObjectsNum}{4\xspace}
\newcommand{\disappearedLinesNum}{3\xspace}
\newcommand{\thicknessOfLineNum}{2\xspace}
\newcommand{\newButtonNum}{2\xspace}
\newcommand{\aNewOutlineAppearNum}{1\xspace}

\newcommand{\categoryLayoutNum}{75}
\newcommand{\layoutShiftNum}{42\xspace}
\newcommand{\specificItemsNum}{10\xspace}
\newcommand{\layoutChangeNum}{12\xspace}
\newcommand{\sizeOfPageHeaderNum}{11\xspace}

\newcommand{\categoryColorNum}{28}
\newcommand{\colorOfObjectsNum}{16\xspace}
\newcommand{\colorOfTextsNum}{8\xspace}
\newcommand{\colorOfLinesNum}{4\xspace}

\newcommand{\categoryTextNum}{18}
\newcommand{\textModificationNum}{5\xspace}
\newcommand{\someTextsDisappearNum}{7\xspace}
\newcommand{\sizeOfTextsNum}{3\xspace}
\newcommand{\someTextsAppearNum}{3\xspace}

\newcommand{\categoryValueNum}{13}
\newcommand{\undefinedValueNum}{8\xspace}
\newcommand{\defaultValueNum}{5\xspace}

\newcommand{\categoryTestNum}{12}
\newcommand{\NoVisualNum}{5\xspace}
\newcommand{\fragileNum}{3\xspace}
\newcommand{\focusNotWorkNum}{2\xspace}
\newcommand{\flackyTestNum}{1\xspace}
\newcommand{\noChangeNum}{1\xspace}

\newcommand{\categoryPictureNum}{8}
\newcommand{\pictureAdjustmentNum}{6\xspace}
\newcommand{\pictureReplacementNum}{2\xspace}

\pgfmathsetmacro{\calcLayout}{(\categoryLayoutNum)/\validVRTclassificationComment*100}
\pgfmathsetmacro{\calcAppearance}{(\categoryAppearanceNum)/\validVRTclassificationComment*100}
\pgfmathsetmacro{\calcColor}{(\categoryColorNum)/\validVRTclassificationComment*100}
\pgfmathsetmacro{\calcText}{(\categoryTextNum)/\validVRTclassificationComment*100}
\pgfmathsetmacro{\calcState}{(\categoryValueNum)/\validVRTclassificationComment*100}
\pgfmathsetmacro{\calcTest}{(\categoryTestNum)/\validVRTclassificationComment*100}
\pgfmathsetmacro{\calcImage}{(\categoryPictureNum)/\validVRTclassificationComment*100}
\definecolor{darkgreen}{rgb}{0, 0.5, 0} 
\definecolor{whitesmoke}{rgb}{0.99, 0.99, 0.99} 

\def\Underline{\setbox0\hbox\bgroup\let\\\endUnderline}
\def\endUnderline{\vphantom{y}\egroup\smash{\underline{\box0}}\\}
\def\|{\verb|}

\newcommand{\ie}{\textit{i.e.,}\xspace}

\newcommand{\etal}{\xspace\textit{et al.}\xspace}

\newcounter{findings_no}

\def\thesubsubsectiondis{\unskip\arabic{subsubsection})}

\usepackage{listings}
\definecolor{backcolour}{rgb}{0.95,0.95,0.92}
\lstdefinelanguage{diff}{
  morecomment=**[f][\color{red}]{-},         
  morecomment=**[f][\color{darkgreen}]{+},       
  moredelim=**[is][\bfseries]{@@}{@@},
}
\definecolor{backcolour}{rgb}{0.95,0.95,0.92}
\lstdefinelanguage{commit}{ 
  breakindent = 0pt,
  numbers=none,
  backgroundcolor=\color{white},
  frame=single,
  xleftmargin=3.5em,
  numbersep=0em,
  xrightmargin=1.5em,
}

\usepackage[many]{tcolorbox}
\tcbuselibrary{listings,breakable}
\definecolor{main}{HTML}{D0D3D4}
\definecolor{sub}{HTML}{D0D3D4}
\newtcolorbox{dbox}{
    left=0pt,right=0pt,top=0pt,bottom=0pt,
    enhanced,
    boxrule=0pt,
    enlarge top by=0pt,
    enlarge bottom by=0pt,
}
\tcbset{
    sharp corners,
    before skip=0pt,
    after skip=0pt
}

\newtcolorbox{rqanswer}{
    sharp corners,
    left=4pt,
    right=4pt,
    top=3pt,
    bottom=3pt,
    boxsep=0pt,
    before skip=3pt,
    after skip=3pt,
}

\def\BibTeX{{\rm B\kern-.05em{\sc i\kern-.025em b}\kern-.08em
    T\kern-.1667em\lower.7ex\hbox{E}\kern-.125emX}}

\begin{document}
\title{What Are Developers Actually Discussing When Visual Regression Tests Fail?}


\author{

\IEEEauthorblockN{
Miku Watanabe\IEEEauthorrefmark{1},
Kosei Horikawa\IEEEauthorrefmark{1},
Brittany Reid\IEEEauthorrefmark{1},
Yutaro Kashiwa\IEEEauthorrefmark{1},
Hajimu Iida\IEEEauthorrefmark{1}}

\IEEEauthorblockA{
\IEEEauthorrefmark{1}{\em Nara Institute of Science and Technology, Japan}}
}

\newcommand\submittedtext{%
 \footnotesize This work has been submitted to the IEEE for possible publication. Copyright may be transferred without notice, after which this version may no longer be accessible.}
\newcommand\submittednotice{%
\begin{tikzpicture}[remember picture,overlay]
\node[anchor=south,yshift=10pt] at (current page.south) {\fbox{\parbox{\dimexpr0.65\textwidth-\fboxsep-\fboxrule\relax}{\submittedtext}}};
\end{tikzpicture}%
}
\maketitle

\begin{abstract}
Visual Regression Tests (VRTs) are widely adopted as a mechanism 
for detecting unintended visual changes in user interfaces. 
By design, VRTs operate on rendered pixel output, and the prevailing 
assumption is that they catch stylistic regressions such as layout 
shifts, color mismatches, and font alterations. We conduct an 
empirical analysis of 307 pull requests (PRs) from 103 GitHub 
repositories that incorporate VRT results via Chromatic, comparing 
them against 299 PRs that contain image attachments but no VRT 
(Visual PRs). Quantitatively, VRT-PRs show no significant acceptance-rate difference, but exhibit a 3.8 times longer median resolution time, 
10 times more discussion comments, and 1.75 to 4.5 times larger 
code changes than Visual PRs. VRT results are typically shared 
around the midpoint of the review process, sustaining ongoing 
discussion rather than serving only as a final check. Through 
a card-sorting analysis of 189 VRT-flagged issues, we identify seven defect categories assigned to the analyzed issues: Layout (39.7\%), Appearance (27.5\%), 
Color (14.8\%), Text (9.5\%), State (6.9\%), Test (6.3\%), and 
Image (4.2\%). 
The three most frequent categories are
stylistic, while approximately 18.5\% of analyzed issues (35/189) involve
non-stylistic origins, including undefined component state (13 cases),
content disappearance (17 cases across multiple categories), and visually
imperceptible regressions (5 cases).
We further document cases in which VRT detected visual regressions
originating from code changes in seemingly unrelated files, exposing
non-local effects that no targeted test would have been written to
catch. These observations indicate that, in addition to its primary
role as a stylistic checker, VRT functions as a secondary detector
of unintended consequences of code changes, with implications for
how VRT should be integrated into the maintenance toolchain.
\end{abstract}







\maketitle

\section{Introduction}\label{sec:introduction}
The quality of a user interface significantly impacts user ratings~\cite{7reason/ui/uninstall}, motivating developers and companies to invest substantial effort in UI improvement~\cite{DBLP:conf/icse/ZhaoCL021}. UI quality has measurable effects on user satisfaction~\cite{DBLP:TOSEM/2021/Chen}\cite{DBLP:journals/sigchi/Jansen98}, engagement~\cite{Durgekar_UI_engage}, and product success~\cite{DBLP:conf/hci/BertramHJLS25}, making it a critical component of modern software engineering practice.

Despite this importance, user interfaces are inherently fragile~\cite{DBLP:journals/corr/abs-2009-01417}. Minor changes in layout, styling, or component behavior can inadvertently degrade user experience~\cite{DBLP:journals/tse/LiuCWHHW23}. This fragility intensifies in large-scale applications where multiple developers work on the same codebase, often introducing inconsistencies or regressions~\cite{DBLP:conf/sigsoft/BirdNMGD11}. As applications evolve through frequent updates and feature additions, maintaining a consistent, high-quality interface becomes increasingly difficult~\cite{DBLP:conf/icse-chase/BaltesD24}.

To address this challenge, many development teams have adopted Visual Regression Tests (VRTs). VRTs automatically capture and compare UI screenshots before and after code changes to detect unintended visual differences. By integrating VRT into the development workflow, teams can identify UI anomalies early in the development cycle, reducing the risk of deploying broken or inconsistent interfaces. Technical blogs report that this proactive approach safeguards user experience while enhancing development efficiency through immediate feedback on visual changes~\cite{chromatic/helping/ui}.

Despite VRT's widespread adoption, only a few studies have examined these tests~\cite{DBLP:conf/apsec/MahajanGPH16}\cite{DBLP:journals/access/PrazinaBCO23}. These studies proposed approaches to detect UI inconsistencies and enhance visual regression testing. To the best of our knowledge, no empirical studies have analyzed VRT activities in practice. Consequently, how visual regression tests contribute to front-end development and what challenges developers encounter when using VRTs remain unclear.

This study clarifies how visual regression tests impact software development and identifies issues detectable through VRTs. We collected the outcomes of VRTs from GitHub by identifying Pull Requests (PRs) linked to Chromatic~\cite{chromaticDocs}, a UI quality management service. Chromatic runs VRT on Storybook~\cite{storybook}, a widely used UI component development environment, and visualizes UI changes via a web application. Analyzing VRT results on Chromatic enables us to examine UI improvement activities in real development contexts.

Our empirical analysis of \prs PRs using Chromatic shows that they are associated with more active discussions and longer merge times. Furthermore, these PRs show a median number of changed files 1.75 times greater than those without VRTs. Additionally, our manual inspection revealed that VRTs not only detect visual changes but also facilitate the identification of unexpected functional regressions.

\vspace{0.5mm}
 {\bf Replication Package: }To facilitate replication studies and future
extensions, the data used in our work is publicly available in the replication package~\cite{Replication}.




\section{Background}
\label{sec:Background}
\subsection{Visual Regression Tests}
Visual Regression Testing (VRT) is a form of software testing that detects unintended visual changes in a user interface (UI) across updates. It ensures the visual presentation of a web or mobile application stays consistent after code changes. A related technique, GUI Testing, simulates user interactions by locating and manipulating UI elements to validate functional behavior, whereas VRT compares screenshots taken before and after changes.

\begin{figure}[t]
    \centering
    \includegraphics[width=1\linewidth]{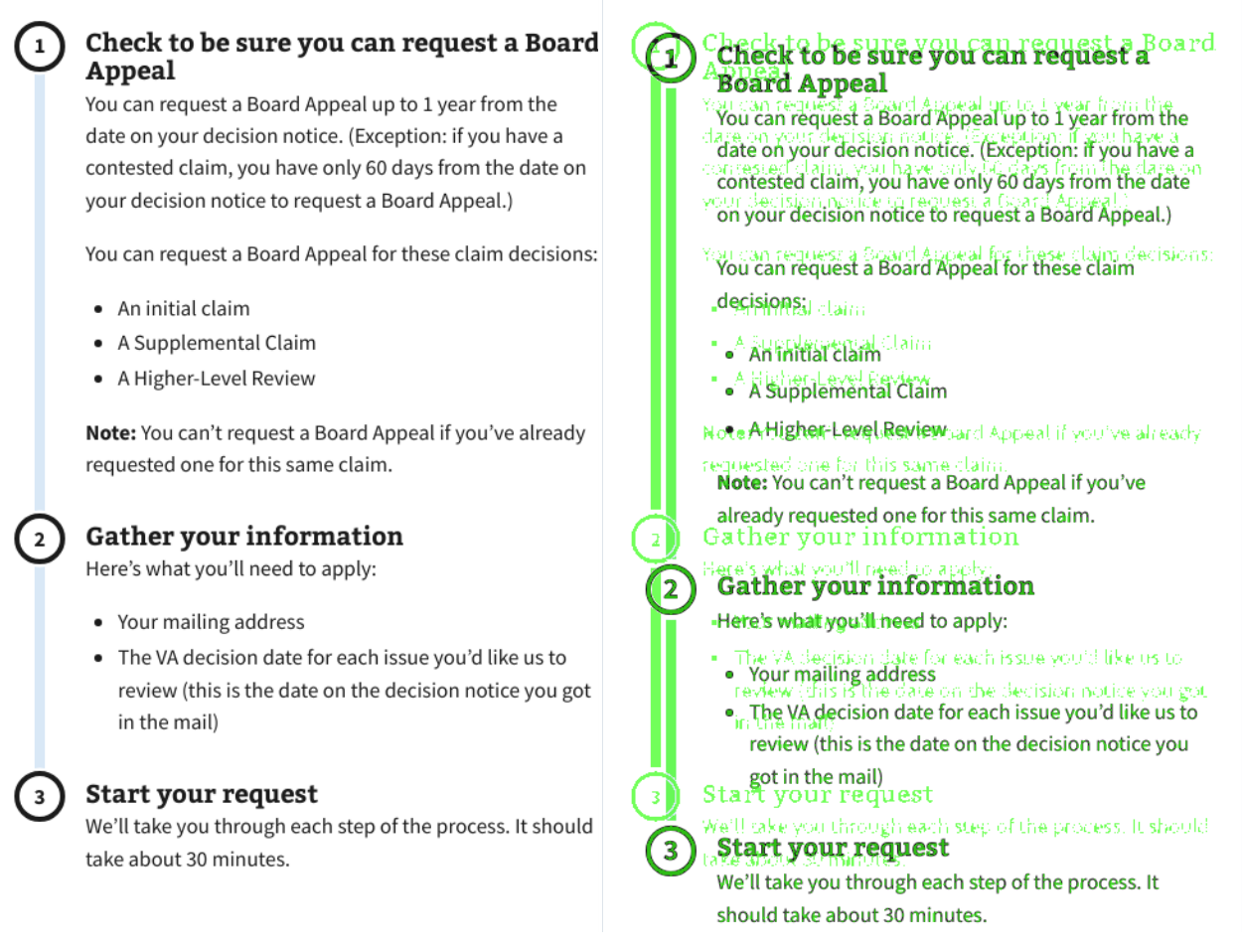}
    \caption[]{Example of UI Regressions Detected by VRT~\cite{ChromaticExample}}
    \label{fig:regression}
\end{figure}

VRT renders the UI in a controlled environment, typically a headless browser, and captures baseline screenshots of the expected state. After code changes, it captures new screenshots of the updated UI and compares them against the baseline using per-pixel or perceptual diffing algorithms. When discrepancies are found, the tool produces a diff image highlighting them. As shown in Fig.~\ref{fig:regression}, VRT tools compare a baseline snapshot (left) with an updated snapshot (right) and flag anomalies such as layout shifts, color mismatches, or text misalignments, commonly highlighted in green.

Developers then review the differences to decide whether they are intentional design changes or regressions. Expected changes are approved and stored as the new baseline; otherwise, the issues are fixed before merging. For example, in one pull request~\cite{ChromaticExpected}, Chromatic flagged a dark mode rendering issue in the sub-navigation component of PictureLayout caused by a copy-paste error that had gone unnoticed, illustrating how VRT can catch subtle defects that might otherwise reach production.

VRT can be plugged into a continuous integration pipeline, enabling automated visual verification of every code change. Despite its growing industry adoption, little is known about the types of visual issues VRT detects, its coverage, and the operational challenges it introduces. This study empirically investigates the impacts of VRT on modern development practices.

\subsection{Related Work}
\label{sec:related work}
GUI testing verifies that the user interface behaves correctly from the user's perspective~\cite{DBLP:conf/icse/LanLLP00L24}\cite{DBLP:conf/icse/RomanoSGYW21}\cite{DBLP:conf/kbse/LanLPL24}. Although it can cost more upfront than manual testing, it tends to be more efficient overall, with higher defect detection rates~\cite{DBLP:journals/ese/AlegrothFR15}. Al\'egroth\etal\cite{DBLP:journals/ese/AlegrothFR15} studied GUI Testing in two safety-critical software companies and found 9 defects, including 5 that manual testing had missed.

Several studies examine snapshot testing~\cite{DBLP:conf/msr/BuiR23}\cite{DBLP:conf/icsm/FujitaK0I23}\cite{DBLP:journals/jss/CruzRV23}, a related concept. Both snapshot and visual regression tests detect unintended UI changes, but snapshot tests target a component's structural output (e.g., generated HTML or JSX) and cannot catch visual issues such as layout shifts or styling errors. Fujita\etal\cite{DBLP:conf/icsm/FujitaK0I23} found that projects using snapshot testing have over 1.6 times as many test cases as those that do not.

Other work has examined UI defects from several angles, including defect classification~\cite{DBLP:journals/infsof/YusopGSV20}, defect location and impact~\cite{DBLP:conf/icst/RobinsonB09}, and user feedback~\cite{DBLP:TOSEM/2021/Chen}. Most closely related to our work, Kuramoto\etal\cite{DBLP:conf/iwpc/KuramotoKKISKU22} studied visual issue reports with images and videos on GitHub, finding that image-containing issues had fewer words than text-only reports while videos showed no word-count difference, and that visual information did not significantly increase discussion activity or speed up resolution.

While these studies characterize GUI functional testing, snapshot testing, or visual issue reports, none has empirically examined VRT activity in PR review contexts; we address this gap by analyzing VRT-linked PRs in their natural development setting.

\section{Data Collection}
We collected PRs from open-source projects that utilize Chromatic, a cloud-based VRT and UI review platform. Chromatic enables developers to detect visual and functional discrepancies in web user interfaces by comparing UI snapshots before and after code changes.

We focused on identifying PRs that include visual regression test results. 
Because many discussions occur in GitHub PR threads, we targeted PRs referencing Chromatic results to examine review discussions and outcomes in context.

To identify such PRs, we searched for those containing URLs linking to Chromatic test result pages, which typically begin with ``\texttt{https://www.chromatic.com/test?}''. We performed this search using the GitHub GraphQL API~\cite{graphql}, configuring the query to exclude bot-generated comments. Our dataset includes PRs created between July 2, 2018 (the release date of Chromatic), and September 30, 2025.

This search yielded \prswithInopen PRs. We excluded \exludepr PRs that remained open (\ie still under development) and removed comments generated by bots that included VRT results. After filtering, we obtained \prs VRT-PRs across \projectsInclosed repositories, with \commentsInclosed comments containing Chromatic links.

We built a comparison dataset of PRs containing image attachments in their descriptions or comments, called Visual PRs~\cite{DBLP:conf/iwpc/KuramotoKKISKU22}. Using the same \projectsInclosed repositories, we searched for \texttt{$<$img$>$} HTML tags in PR descriptions or comments during the same period (July 2, 2018, to September 30, 2025). To ensure comparable representation, we applied stratified random sampling: for each repository with VRT-PRs, we randomly sampled an equal number of Visual PRs from the same period, excluding any VRT-PRs. This yielded \visualPrs Visual PRs compared to \prs VRT-PRs. The difference reflects the limited availability of Visual PRs in these repositories.

\section{Results}

\subsection*{\rqA}


\motivation The outputs of VRTs provide concrete evidence of UI changes, which could either facilitate quicker approvals through clearer communication or extend review time by revealing issues requiring discussion.
Kuramoto\etal\cite{DBLP:journals/ese/KuramotoWKKKU24} found that issues highlighted by images tend to be resolved more quickly. Since VRT provides more structured visual feedback than screenshots, it may similarly affect review resolution.


\approach To investigate whether VRT presence is associated with PR acceptance and resolution time, we compare VRT-PRs and Visual PRs in terms of their acceptance rates and merge times.
We define the acceptance rate as the proportion of merged PRs among all considered PRs (\ie both merged and closed without merging). Merge time represents the number of days between a PR's creation and merge date; we calculate this metric only for merged PRs.
To assess statistical significance, we apply the Chi-square test to compare acceptance rates and the log-rank test to compare merge time distributions between the two groups ($\alpha = 0.01$). 



\results Regarding acceptance rates, we found that \acceptRateVrt\% of VRT-PRs (\mergedVrtPR{}/\prs PRs) and \acceptRateVisual\% of Visual PRs (\mergeVisualPR{}/\visualPrs)  were merged. 
Although the acceptance rate of VRT-PRs was 5.3\% points higher than that of Visual PRs, the difference was not statistically significant.

\begin{table}[!th] 
  \centering 
  \caption{Resolution time (days) for merged PRs} 
  \begin{tabular}{@{}lrrrrrrr@{}} 
    \toprule
    Project & \#PRs&Max& Mean & Median& Std      \\
    \midrule
    Visual-PRs &\mergeVisualPR{}&146.0 & 6.8 &  \textbf{\timeMergeMedianVisual} & 16.9 \\
    VRT-PRs    &\mergedVrtPR{}&173.2 & 11.2 & \textbf{\timeMergeMedianVrt} & 20.5 \\
    \bottomrule
  \end{tabular}
    \label{tab:time_to_merge_stats_transposed} 
\end{table}

Table~\ref{tab:time_to_merge_stats_transposed} summarizes the resolution times of VRT-PRs and Visual PRs. We found that the median resolution time for VRT-PRs is \num{\calcTimeDiff} times longer than that of Visual PRs. To assess the statistical significance of this difference, we conducted a log-rank test ($\alpha = 0.01$), which revealed a statistically significant difference in resolution times between the two groups. This diverges from Kuramoto\etal\cite{DBLP:conf/iwpc/KuramotoKKISKU22}, 
who reported that visual issue reports did not significantly 
affect resolution time; VRT-PRs are instead resolved notably 
more slowly, suggesting VRT outputs trigger substantive review 
rather than routine confirmation.

\begin{rqanswer}
\textbf{Answer to RQ1.}{
\textit{The 5.3-point acceptance-rate difference was not significant, whereas VRT-PRs took significantly longer to be resolved.}
}
\end{rqanswer}

\subsection*{\rqB}
\motivation Spadini\etal\cite{DBLP:conf/icse/SpadiniASBB18} observed that test files are nearly twice as likely to receive less discussion during code review when reviewed alongside production files. Since VRTs are a form of testing, they may exhibit similar patterns. However, VRTs differ from traditional test code (which focuses on functional aspects) by providing visual feedback on UI changes, potentially affecting how reviewers engage.

\approach We compare the number of comments between the merged VRT-PRs and Visual PRs (\mergedVrtPR{} and \mergeVisualPR{} PRs, respectively). Specifically, we measure the total number of comments associated with each PR, including both review and discussion comments. To determine whether the differences are statistically significant, we apply the Mann-Whitney U test.

Additionally, we examine when comments containing links to VRT results appear in the review process to understand how VRT outputs are used during reviews. We normalize the comment index by the total number of comments in each PR to assess the relative timing of these VRT-related comments.


\results We compared the number of comments in merged VRT-PRs 
and Visual PRs. The median number of comments in VRT-PRs was 
\commentMedianVrt{},  \calcCommentDiff{} 
times as many as in Visual PRs (\commentMedianVisual{} comment). 
This difference is statistically significant ($p < 0.01$), and 
the effect size (r = \commentEffect) indicates a large practical 
difference. This contrasts with two prior observations: 
Spadini\etal\cite{DBLP:conf/icse/SpadiniASBB18} found that test 
files receive less discussion than production code, and 
Kuramoto\etal\cite{DBLP:conf/iwpc/KuramotoKKISKU22} reported that 
visual information did not significantly increase discussion 
activity. VRT-PRs behave differently from both, eliciting an 
order of magnitude more comments.

Looking into the timing of comments that include links to VRT 
results, these links typically appear after 50.0\% of the total 
comments have already been made (median). This suggests that 
VRT results are referenced not only during final validation but 
also throughout the discussion process, potentially facilitating 
earlier and more informed collaboration.

\begin{rqanswer}
\textbf{Answer to RQ2.} \textit{VRT-PRs receive significantly more comments than Visual PRs, indicating more active discussion. Additionally, VRT results are shared in the middle of the reviewing process rather than only at completion.}
\end{rqanswer}

\subsection*{\rqC}
\motivation Many studies~\cite{DBLP:conf/icse/ZhongS15}\cite{DBLP:conf/icse/Hassan09} use metrics such as the number of commits and the size of code changes to characterize the complexity of fixing issues.
Since VRTs often detect subtle visual differences across multiple components, they may require broader or more intricate modifications.

\approach We examine whether VRT-PRs involve more extensive code changes compared to Visual PRs. We compare merged VRT-PRs (\ie \validVRTprs PRs) and merged Visual PRs (\ie \visualMergedPrs PRs) using metrics commonly used to characterize the scale and complexity of code modifications: number of commits, changed files, and lines added and deleted. 

\begin{table}[!bt]
  \centering
  \caption{Effectiveness Measurement}
  \label{tab:comparison}
  \begin{tabular}{lrrrr} 
    \toprule
    & \multicolumn{2}{c}{\textbf{Median}} & & \\
    \cmidrule(lr){2-3}
    \textbf{Metric} & \textbf{Visual-PR} & \textbf{VRT-PR} & \textbf{p-value} & \textbf{Effect Size}\\
    \midrule
    Commits       &  \commitsVisual{} & \commitsVrt{} & \textbf{$<$ 0.001} & \commitCiff(M)\\
    Changed files & \changeFileVisual{} & \changeFileVrt{} & \textbf{$<$ 0.001} & \changeFileCiff(S) \\
    Added lines   & \addedLineVisual & \addedLineVrt & \textbf{$<$ 0.001}            & \addedCiff(S)\\
    Deleted lines & \deletedLineVisual  & \deletedLineVrt & \textbf{$<$ 0.001}  & \deleteCiff(M)\\
    \bottomrule
  \end{tabular}
\vspace{-5mm}
\end{table}

\results Table~\ref{tab:comparison} presents the distribution of code change metrics between Visual-PRs and VRT-PRs.  
The median number of commits in Visual-PRs is \commitsVisual{}, compared to \commitsVrt{} in VRT-PRs. This is 2.3 times larger.
Other metrics also show more extensive changes in VRT-PRs. The median number of changed files, added lines, and deleted lines ranges from 1.75 to 4.5 times higher than in Visual-PRs.
All differences are statistically significant ($p < 0.001$).  
The corresponding effect sizes indicate medium to small practical differences: r = \commitCiff for commits (medium), \changeFileCiff for changed files (small), \addedCiff for added lines (small), and \deleteCiff for deleted lines (medium).



\begin{rqanswer}
\textbf{Answer to RQ3.}{
\textit{VRT-PRs exhibit about twice the median number of commits compared to Visual-PRs, along with 1.75 to 4.5 times more changed files and lines of code.}
}
\end{rqanswer}

\subsection*{\rqD}
\motivation Kuramoto\etal\cite{DBLP:journals/ese/KuramotoWKKKU24} categorized issues reported with images or videos, finding that visual evidence is frequently used to illustrate program behavior or UI layout. While VRTs similarly generate visual artifacts that developers reference during code review, the specific types of issues detected by VRTs remain unclear.

\approach We used a card sorting approach~\cite{spencer2009card} to classify the \validVRTcomments{} issues found in merged VRT-PRs (\ie 344 of the 373 Chromatic-link comments). Two authors independently inspected review comments containing Chromatic links and examined the corresponding Chromatic outputs. They annotated the issues identified by VRTs to capture their intent, with each case allowed to receive multiple labels. The authors then discussed and reconciled their labels.
The initial independent classification by the first and second authors achieved 71.8\% label-level agreement, with disagreements on 118 of 419 labels. For disputed cases, a third author reviewed the conflicting labels and recommended resolutions. These recommendations were discussed with the original two authors until complete agreement was reached. The three authors have 8-17 years of programming experience.

It is worth noting that some PRs were written in non-English languages. Given the limited dataset size and the importance of maintaining diversity, we used translation services to interpret the content rather than excluding these PRs.

\results During manual inspection, we initially created 32 labels and then grouped similar ones into 7 categories comprising 29 labels in total, as summarized in Table~\ref{table:label}. Of the \validVRTcomments{} review issues inspected, we excluded \invalidcomment invalid cases: duplicate links (\duplicate issues), inaccessible pages (\inaccessible issues), Chromatic errors (\chromaticerror issues), new stories (\newstory issues), and those lacking sufficient context (\lackofcontext issues), leaving \validVRTclassificationComment~issues for analysis. 
The category distribution shows VRT's primary role: the three most frequent categories (Layout, Appearance, Color) all concern visual styling. However, 35 of the 189 issues (18.5\%) involve non-stylistic origins: undefined component state (13, the State category), content disappearance (17: \texttt{Contents}, \texttt{Box}, and \texttt{Text Disappeared}), and visually imperceptible regressions (5, \texttt{No or Few Visual Differences Detected}). These cases suggest a secondary detection role.

\begin{table}[t] 
  \centering 
  \caption{Categories of issues that developers discuss}
  \label{table:label} 
  \begin{tabularx}{\linewidth}{lX} 
    \toprule
    \textbf{Category} & \textbf{Labels} \\
    \midrule

    \multirow{2}{*}{\categoryLayout (\categoryLayoutNum)}& \layoutShift(\layoutShiftNum), \layoutChange(\layoutChangeNum), \sizeOfPageHeader(\sizeOfPageHeaderNum),  \specificItems(\specificItemsNum)  \\
    \addlinespace %
    \multirow{4}{*}{\begin{tabular}{@{}l@{}}\categoryAppearance \\ \multicolumn{1}{l}{(\categoryAppearanceNum)}\end{tabular}} & \sizeOfObjects (\sizeOfObjectsNum), \newContents (\newContentsNum), \disappearedContents (\disappearedContentsNum), \shapeOfObjects (\shapeOfObjectsNum), \disappearedObjects (\disappearedObjectsNum), \disappearedLines (\disappearedLinesNum), \thicknessOfLine (\thicknessOfLineNum), \newButton (\newButtonNum), \aNewOutlineAppear (\aNewOutlineAppearNum)  \\ %
    \addlinespace 
    \multirow{2}{*}{\categoryColor (\categoryColorNum)}& \colorOfObjects (\colorOfObjectsNum), \colorOfTexts (\colorOfTextsNum), \colorOfLines (\colorOfLinesNum) \\
    \addlinespace
    \multirow{2}{*}{\categoryText (\categoryTextNum)} & \someTextsDisappear (\someTextsDisappearNum), \textModification (\textModificationNum), \sizeOfTexts (\sizeOfTextsNum), \someTextsAppear (\someTextsAppearNum) \\
    \addlinespace
    \categoryValue (\categoryValueNum)& \undefinedValue (\undefinedValueNum), \defaultValue (\defaultValueNum) \\
    \addlinespace
    \multirow{2}{*}{\categoryTest (\categoryTestNum)} & \NoVisual (\NoVisualNum), \fragile (\fragileNum), \focusNotWork (\focusNotWorkNum), \flackyTest (\flackyTestNum), \noChange  (\noChangeNum) \\ 
    \addlinespace
    \categoryPicture (\categoryPictureNum) & \pictureAdjustment (\pictureAdjustmentNum), \pictureReplacement (\pictureReplacementNum) \\
    \bottomrule

  \end{tabularx}
\vspace{-5mm}
\end{table}
\noindent\textbf{\categoryLayout:}
This category accounted for \num{\calcLayout}\% of issues (\categoryLayoutNum{}/\validVRTclassificationComment{}) and was the most frequently observed type of visual regression. These VRT detections involved inconsistencies in the placement, spacing, and alignment of UI elements. The most prominent subtype was \texttt{\layoutShift}, in which items were unintentionally repositioned, accounting for \layoutShiftNum cases.
In one representative case, a developer identified a flaw in the baseline itself: an unintended widening of the left column. They argued that the updated rendering reflected the correct grid layout~\cite{LayoutChromatic}.
This detection prompted a broader discussion on layout principles, including the guideline that \textit{``the grid should not break due to long section titles; instead, titles should wrap to multiple lines if necessary.''} This case shows how VRT can reveal not only regressions but also misalignments in the baseline, triggering active discussion~\cite{LayoutGithub}.

\noindent\textbf{\categoryAppearance:}
This category accounts for \num{\calcAppearance}\% of issues (\categoryAppearanceNum{}/\validVRTclassificationComment{}) and represents the second most frequently observed type of regression related to the visual appearance of the UI. Specifically, \texttt{\sizeOfObjects} (\sizeOfObjectsNum cases), \texttt{\newContents} (\newContentsNum cases), and \texttt{\disappearedContents} (\disappearedContentsNum cases) are the most common. 
In one case~\cite{AppearanceChromatic}, when developers improved the basic layer structure of the UI, the impact extended beyond expectations, causing a panel to disappear from a separate component. VRT detected this change, and the discussions covered specific countermeasures, release timing adjustments, and a support policy for the older component.


\noindent\textbf{\categoryColor:}
This category accounted for \num{\calcColor}{}\% of issues (\categoryColorNum{}/\validVRTclassificationComment{}) and includes regressions related to the color scheme of UI elements, such as text color, background color, and line color. The most common subtypes were \texttt{\colorOfObjects} (\colorOfObjectsNum cases) and \texttt{\colorOfTexts} (\colorOfTextsNum cases).

A representative example is shown in a PR~\cite{ColorGithub} where VRT detected an unintended color change. The issue was originally caused by a previous PR that modified unrelated files and included no discussion of visual changes.
Interestingly, the PR that introduced the regression affected different parts of the codebase and included no discussion of VRT results, highlighting two key insights: (1) visual regressions can propagate indirectly across seemingly unrelated changes, and (2) VRT is effective in surfacing such issues early, even when developers are unaware of their visual impact.


\noindent\textbf{\categoryText:}
This category accounted for \num{\calcText}\% of issues (\categoryTextNum{}/\validVRTclassificationComment{}) and includes regressions related to text content and rendering. The most common subtypes were \texttt{\someTextsDisappear} (\someTextsDisappearNum~cases). 
In a representative case, a developer was notified by the VRT that a sentence had unexpectedly and unintentionally disappeared from the user interface~\cite{TextGithub}.


\noindent\textbf{\categoryValue:}
This category accounted for \num{\calcState}\% of issues (\categoryValueNum{}/\validVRTclassificationComment{}) and includes regressions caused by incorrect or missing property settings or bugs in the UI. The most common subtype was \texttt{\undefinedValue} (\undefinedValueNum~cases), where UI elements rendered unexpected content due to misconfigured properties. 
For example, developers identified a defect in which a variable was not correctly assigned, resulting in an undefined value being displayed in the UI~\cite{StateGithub}.



\noindent\textbf{\categoryTest:}
This category accounted for \num{\calcTest}\% of issues (\categoryTestNum{}/\validVRTclassificationComment{}) and includes instances where no obvious visual differences were initially apparent, yet VRT still flagged changes. A representative sub-category is \texttt{\NoVisual} (\NoVisualNum cases).
In one such case, although VRT detected a visual difference, the developers were unable to perceive any noticeable change in the rendered UI~\cite{TestGithub}.

\noindent\textbf{\categoryPicture:}
This category accounted for \num{\calcImage}\% of issues (\categoryPictureNum{}/\validVRTclassificationComment{}) and includes regressions related to images and pictures, with the most common subtypes being \texttt{\pictureAdjustment} (\pictureAdjustmentNum~cases) and \texttt{\pictureReplacement} (\pictureReplacementNum cases). In one representative case, VRT detected visual discrepancies in an image component, where the aspect ratio had changed and the credit attribution was missing from the rendered output~\cite{ImageGithub}.



\begin{rqanswer}
{\textbf{Answer to RQ4.}}{
\textit{Of the seven categories identified, several (notably State, Test, 
and parts of Appearance and Text) consist of defects whose visual symptoms 
originate in functional, propagation-related, or sub-perceptual causes 
rather than stylistic ones, indicating that VRT functions as a detector of 
unintended consequences of code changes rather than a purely visual checker.}
}
\end{rqanswer}


\section{Future Research Directions}
This section discusses our findings and how they can inform the direction of future studies.

\noindent\textbf{Distinguishing intentional changes from unintended regressions. }
RQ4 indicates that VRTs detect a diverse range of issue types that differ in nature. Additionally, RQ1 shows that issues involving VRT take longer to be fixed, and RQ2 revealed that VRT results are discussed during the review process.
While these findings suggest that issues involving VRT are not trivial, they need evaluation based on whether they represent intentional changes or unintended regressions.
For instance, most \texttt{Layout shifted} issues in the Layout category tend to occur unexpectedly, without developer intent. In contrast, \texttt{New Contents} in the Appearance category may generally reflect deliberate changes introduced by developers. Future work should identify whether each failure detected by VRT was expected or unintended and develop methods to make this determination automatically, thereby reducing investigation efforts.

\noindent\textbf{Understanding the cause of visual regressions and their fixes. }
RQ3 reveals that the underlying code changes associated with VRT failures are typically more substantial, implying that larger or more complex modifications are more likely to introduce visual regressions.
Future research should study the causes of unintentional test failures included in changes. Insights from these investigations could inform best practices for preventing visual regressions and support the development of automated repair techniques~\cite{DBLP:conf/issre/DurieuxHM18}\cite{DBLP:conf/icse/OcarizaPM14}.

\section{Threats to Validity}\label{sec:threats_to_validity}
We acknowledge several limitations that may affect the validity of our findings. Below, we discuss potential threats across three dimensions.

\noindent\noindent{\bf Threats to internal validity:} This study relies on human annotations, which can be subjective. To mitigate the potential bias, we examined the results of VRTs independently and discussed conflicting annotations until full agreement. 

\noindent{\bf Threats to construct validity:} 
Our study reports associations but does not establish causality. VRTs may be disproportionately applied to PRs with substantial UI modifications, which inherently require more discussion and larger changes. The observed differences may partly reflect the complexity of underlying changes rather than the effect of VRT itself. Disentangling these effects would require matched-pair analyses comparing PRs of similar scope with and without VRT.

\noindent{\bf Threats to external validity:}
This study examines only \prs PRs that contain links to the results of VRTs. Although many projects today employ VRTs, it remains challenging to reliably identify both the test results and the associated discussions within PRs. To mitigate this issue, we included \textit{all instances} we were able to retrieve at the time of data collection. 



\section{Conclusions}\label{sec:conclusion}
This study analyzed \prs{} pull requests from GitHub repositories 
employing VRTs. Compared to Visual PRs, VRT-PRs show no significant acceptance-rate difference, but exhibit a median resolution time 3.8 times longer, 10 times
more discussion comments (median 10 vs.\ 1), and 1.75 to 4.5 times 
larger code changes. These differences indicate that VRT-flagged issues 
are non-trivial and that VRT supports collaborative review; VRT results 
are typically shared around the midpoint of review, sustaining ongoing 
discussion rather than serving only as a final check.

Manual inspection of \validVRTclassificationComment{} issues yielded 
seven categories, 
with Layout (39.7\%), Appearance (27.5\%), and Color (14.8\%) the most 
frequent. Several of these capture functional regressions such as undefined property values 
or disappearing components, indicating that VRT detection extends beyond 
stylistic verification.



\section*{Acknowledgment} 
We gratefully acknowledge the financial support of JSPS KAKENHI grants (JP24K02921, JP25K03100, JP26K23818), as well as JST BOOST (JPMJBS2423), ASPIRE grant (JPMJAP2415), and CREST grant (JPMJCR23M1, JPMJCR26X7).


\balance
\bibliographystyle{IEEEtran} 
\bibliography{references.bib}



\end{document}